\documentclass[10pt,letterpaper]{article}
\usepackage[top=0.85in,left=2.75in,footskip=0.75in]{geometry}

\usepackage{amsmath,amssymb}

\usepackage{changepage}

\usepackage{textcomp,marvosym}
\usepackage{cite}

\usepackage{nameref,hyperref}

\usepackage[right]{lineno}

\usepackage{microtype}
\DisableLigatures[f]{encoding = *, family = * }

\usepackage[table]{xcolor}

\usepackage{array}

\newcolumntype{+}{!{\vrule width 2pt}}

\newlength\savedwidth

\raggedright
\usepackage[aboveskip=1pt,labelfont=bf,labelsep=period,justification=raggedright,singlelinecheck=off]{caption}

\makeatletter
\renewcommand{\@biblabel}[1]{\quad#1.}
\makeatother

\usepackage{lastpage,fancyhdr,graphicx}
\usepackage{epstopdf}
\fancyheadoffset[L]{2.25in}
\fancyfootoffset[L]{2.25in}
\usepackage{here} 
\graphicspath{{./figures/}}
\usepackage{ulem}
\usepackage{bm}

\newif\ifshowred
\showredfalse

\newif\ifshowfig
\showfigtrue

\newcommand{\red}[1]{\ifshowred\textcolor{red}{#1}\else#1\fi}

\begin{document}

\vspace*{0.2in}

\begin{flushleft}
{\Large
\textbf\newline{Population dynamics of generalist and specialist strategies under feast-famine cycles} 
}
\newline 
\\
Rintaro Niimi\textsuperscript{1},
Chikara Furusawa\textsuperscript{2,3*},
Yusuke Himeoka\textsuperscript{2*}
\\
\bigskip
\textbf{1} Department of Biological Sciences, Graduate School of Science, The University of Tokyo, Bunkyo-ku, Tokyo, Japan
\\
\textbf{2} Universal Biology Institute, Graduate School of Science, The University of Tokyo, Bunkyo-ku, Tokyo, Japan
\\
\textbf{3} Center for Biosystems Dynamics Research, RIKEN, Chuo-ku, Kobe, Japan
\bigskip

* furusawa@ubi.s.u-tokyo.ac.jp (CF), yhimeoka@ubi.s.u-tokyo.ac.jp (YH)

\end{flushleft}

\section*{Abstract}
Microbial populations exhibit a broad spectrum of nutrient utilization strategies, 
ranging from those utilizing diverse nutrients, called ``generalists,'' 
to those highly adapted to specific nutrients, called ``specialists.'' Identifying the conditions for the diversification of nutrient utilization strategies is one of the central questions in ecology.
Previous theoretical studies have shown that trade-offs among different resource utilization functions in which cells cannot utilize broad types of substrates at nearly optimal efficiency are crucial for the emergence of diverse strategies. 
Additionally, in natural settings, nutrient availability often fluctuates over time, imposing another trade-off on the cells; cells that grow rapidly under nutrient-rich conditions tend to have a  higher death rate under nutrient-poor conditions, leading to a growth-death trade-off.
This additional trade-off can contribute to the emergence of diverse strategies.
Here, we introduce a mathematical model that simultaneously incorporates the resource-use trade-off and the growth-death trade-off. Nutrient supply was modeled as discrete stochastic events, mimicking temporal changes in nutrient availability.
We show that the phenotype with a higher ratio of growth rate to death rate dominates the population; that is, the strength of the growth-death trade-off plays a crucial role in the emergence of distinct strategies.
We also found that a sparse and uncertain nutrient supply favors specialists, increasing their temporally averaged abundance.
Our findings highlight the crucial role of temporal environmental variation and the resulting growth-death trade-off in driving diversification of microbial nutrient utilization strategies.

\section*{Author summary}
Microbial strategies for choosing the nutrients to utilize are highly diverse.
In our study, we explored why some microbes become generalists able to use a variety of nutrients, whereas others specialize in only a few nutrients.
These strategies often result from metabolic trade-offs during nutrient processing.
However, in fluctuating environmental conditions, it is important to consider additional trade-offs; under nutrient-rich conditions, they can grow quickly,
but this often comes at the cost of being more vulnerable when nutrients are scarce.
To investigate how these trade-offs affect the emergence of generalists and specialists, 
we developed a simple model of a feast-famine cycle that mimics the natural cycles of plenty and scarcity of nutrients.
Our work shows that the trade-off between growth and death rates is crucial for determining whether a generalist or specialist strategy is dominant.
We highlight the importance of trade-offs in response to environmental changes in shaping microbial adaptive strategies.


\section*{Introduction}
Microbes are the most abundant and diverse organisms on Earth \cite{larsenInordinateFondnessMultiplied2017}, inhabit a wide range of environments and form the fundamental basis of ecosystems \cite{whitmanProkaryotesUnseenMajority1998}. 
Contrary to these observations, classical theories argue that increasing species diversity within an ecosystem tends to reduce stability, making the coexistence of species in a single habitat difficult
\cite{
  mayWillLargeComplex1972,stoneFeasibilityStabilityLarge2018
}.
Resolving such discrepancies between theories and observations has been a central question in the field of mathematical ecology  
\cite{chessonMechanismsMaintenanceSpecies2000a}.

\red{
One key factor for coexistence is differentiation in nutrient utilization \cite{chessonMechanismsMaintenanceSpecies2000a}. By partitioning resource use, species reduce overlaps in the consumption of limiting nutrients. This weakens direct competition and makes it easier for multiple species to coexist within the same habitat
\cite{
  brochetNichePartitioningFacilitates2021,
  yuLowlevelResourcePartitioning2024
}.
}

\red{What generates and maintains the differentiation of nutrient utilization strategies across species? A major explanation is the presence of physiological trade-offs\cite{kneitelTradeoffsCommunityEcology2004}.}
Laboratory experiments have documented trade-offs in microbes’ abilities to use different nutrients, including fructose versus galactose 
\cite{ekkersTradeOffsPredictedMetabolic2022a}, silicon versus\ phosphorus 
\cite{kilhamDynamicsLakeMichigan1986}, and nitrogen versus phosphorus 
\cite{edwardsEvidenceThreewayTradeoff2011}.
When cells enhance their ability to use one nutrient, the utilization of other nutrients is constrained, which promotes differentiation of resource use among species.

\red{
Mechanistically, the trade-off can arise from multiple metabolic constraints.
Limited proteome capacity forces cells to allocate protein resources across pathways
\cite{scottInterdependenceCellGrowth2010, 
huiQuantitativeProteomicAnalysis2015,
caetanoEvolutionDiversityMetabolic2021}.
Thermodynamic constraints can also contribute. For example, glycolysis and gluconeogenesis share reactions but operate in opposite directions, making the simultaneous use of glycolytic and gluconeogenic carbon sources infeasible \cite{schinkGlycolysisGluconeogenesisSpecialization2022,ekkersTradeOffsPredictedMetabolic2022a}.
}

The presence of such inherent trade-offs compels microbes to choose which nutrients to use, and may lead to a variety of nutrient utilization strategies.
The two extremes of nutrient utilization strategies are \textit{generalists} and \textit{specialists}. 
Generalists can adapt to diverse conditions, but the utilization speed and/or efficiency of nutrients are typically lower than that of other strategies. 
Specialists use a narrow range of nutrients but achieve high growth rates in their adapted environments
\cite{futuymaEvolutionEcologicalSpecialization1988,sextonEvolutionEcologicalNiche2017}.

Temporal variability of environmental conditions is another key aspect of the emergence of diverse ecosystems \cite{tachikawaFluctuationInducesEvolutionary2008, cooperExperimentalEvolutionColi2010, kremerCoexistenceVariableEnvironment2013, kassenExperimentalEvolutionSpecialists2002, sachdevaTuningEnvironmentalTimescales2020}.
In natural environments, such as lakes, ocean surfaces, and soils, nutrients arrive as discrete and stochastic pulses rather than as a continuous input. Each pulse creates a brief feast phase followed by famine once nutrients are depleted; this alternation is known as the feast-famine cycle \cite{srinivasanCyclesFamineFeast1998, simonMicrobialEcologyOrganic2002, merrittFrequencyAmplitudeDependentMicrobial2018, himeokaDynamicsBacterialPopulations2020}.

Under feast-famine conditions, cells must allocate resources between rapid growth and stress protection.
In {\it E. coli}, for example, a key regulator is {\it rpoS}, which encodes $\sigma^{38}$ (RpoS), the stress-response sigma factor. Deletion of {\it rpoS} ($\Delta$\textit{rpoS}) increases growth rate under nutrient-rich conditions but compromises stress tolerance under starvation \cite{notley-mcrobbRpoSMutationsLoss2002}. 
Because proteome remodeling upon sudden starvation is limited, particularly in bacteria with relatively low protease activity \cite{mauriziProteasesProteinDegradation1992}, cells often maintain basal expression of stress-response genes in advance, improving preparedness at a growth cost \cite{patangeEscherichiaColiCan2018}. 
This leads to a growth-death trade-off across feast and famine. Indeed, a study by Biselli et al. \cite{biselliSlowerGrowthEscherichia2020} showed that there is a positive correlation between growth rate under nutrient-rich conditions and death rate in subsequent carbon starvation.

\red{
Taken together, microbes experiencing a feast-famine cycle may face not only a constraint between the breadth of utilizable nutrients and growth rate, 
but also a constraint between the growth rate during the feast phase and survival rate during the famine phase.
Hereafter, we term the first and second trade-offs the ``resource-use trade-off'' and the ``growth-death trade-off,'' respectively.
}

\red{
How do these two trade-offs together drive the diversity of nutrient utilization strategies?
To date, the resource-use trade-off has been actively investigated and is considered one of the key factors in determining whether generalist- or specialist-like strategies are favored in a given environment.
According to several theoretical studies, when adaptation to one resource substantially impairs performance on others (i.e., the trade-off is strong), specialist strategies tend to be favored; conversely, when the trade-off is weak, generalists tend to be favored
\cite{egasEvolutionRestrictsCoexistence2004,ehrlichTraitFitnessRelationships2017,caetanoEvolutionDiversityMetabolic2021}. 
}

\red{
In such frameworks focusing on the resource-use trade-off, the optimal nutrient utilization strategy is primarily determined by the strength of the trade-off represented by the functional form of the trade-off. 
Accordingly, environmental conditions such as nutrient amount or supply frequency have a limited influence on the emergence of distinct nutrient utilization strategies.
}

\red{
On the other hand, theoretical studies of microbial competition in temporally varying environments (e.g., serial-dilution cycles) have shown that environmental parameters can strongly shape which strategies are favored and how diverse communities become \cite{manhartGrowthTradeoffsProduce2018, erezNutrientLevelsTradeoffs2020, fridmanFinescaleDiversityMicrobial2022, bloxhamDiauxicLagsExplain2022, bloxhamBiodiversityEnhancedSequential2024}. 
In particular, the nutrient amount per pulse and the supply frequency can be key determinants. 
More broadly, temporal variation in resource availability and supply regimes is known to influence community structure and ecosystem processes across diverse ecosystems, including tropical systems \cite{vitousekNUTRIENTCYCLINGMOIST1986}. 
Thus, optimal strategies are not necessarily determined solely by the functional form of a resource-use trade-off. 
While prior studies have largely focused on feast-phase growth dynamics, here we highlight an additional constraint arising from famine-phase survival, captured by a growth-death trade-off, and ask how it interacts with the resource-use trade-off under stochastic feast-famine cycles.
}

To determine how the resource-use trade-off and the growth-death trade-off synergistically affect nutrient utilization strategies, 
we modeled multiple phenotypes competing for nutrients under feast-famine cycles.
In this model, nutrients are supplied by discrete and stochastic events that initiate the feast period.
Once the feast period begins, cells consume nutrients to grow, leading to famine.
We introduce two types of trade-offs for the microbes: if a phenotype utilizes a wide variety of nutrients, it cannot show growth as rapid as another phenotype that grows only on a limited number of nutrient sources.
Additionally, if a phenotype has a higher growth rate during the feast period, it shows a higher death rate under famine conditions.

Using this model, we showed that either a generalist or specialist strategy can be dominant in the population, depending on environmental conditions. 
The fittest strategy shifts from specialists to generalists once the average growth rate exceeds a critical value.
We also found that specialists gain a relative advantage over generalists as supply becomes sparser and uncertain.
Our results provide insights into a mathematical description of the diversification of resource use strategies in temporally varying environments.

\section*{Models}
\begin{figure}[h!]
  \ifshowfig\includegraphics[]{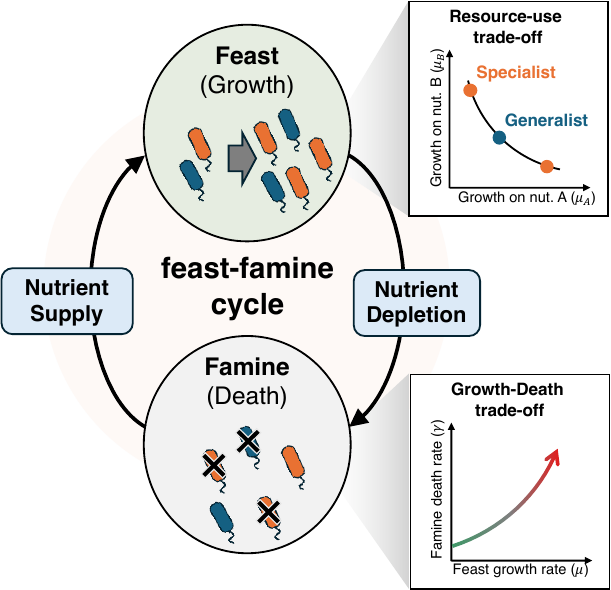}\fi
  \caption{\red{{\bf Schematic illustration of the model.}}
    \red{The central cycle depicts the environmental dynamics where the population alternately experiences the feast phase and the famine phase. Discrete nutrient supply events trigger the transition to the feast phase, whereas nutrient consumption by cells leads to resource depletion, driving the transition to the famine phase. The panels on the right illustrate the two key physiological constraints assumed in the model: (Top) The resource-use trade-off determines the feasible growth rates on distinct nutrients, which defines the spectrum between specialists (orange) and generalists (blue). (Bottom) The growth-death trade-off imposes a cost on fast growth, assuming that phenotypes with higher maximum growth rates ($\mu$) suffer from higher death rates ($\gamma$) during the famine phase.}
    }
    \label{fig:schematicFigure}
\end{figure}

In this study, we investigated the influence of trade-offs on the emergence of multiple nutrient utilization strategies
in a fluctuating environment. 
To this end, we considered a population dynamics model for microbes in the feast-famine cycle \red{(Fig~\ref{fig:schematicFigure})}. 
In the following, we use the term “phenotype” instead of species because different phenotypes differ only in their nutrient utilization strategy within the resolution of the present model.

The model consists of $N$ phenotypes. Nutrients are supplied to the environment as discrete stochastic events, whereas the population growth of the phenotypes is modeled using continuous-time deterministic ordinary differential equations. 
Here, we consider that there are multiple types of nutrients, including different carbon sources. 
There are $E$ types of nutrient sources and a single type of nutrient is supplied by a single supply event. 
As will be seen, the supplied nutrient is almost always consumed much earlier than the next supply event, and thus, at most one type of nutrient is present in the system simultaneously. 
Hence, the type of nutrient supplied at the latest supply event sets the environmental condition;
thus, we represent the environmental condition by the type of nutrient at the latest supply.
Each nutrient supply event initiates a feast period during which cells consume the supplied nutrients and proliferate.
\red{
At the proliferation event of an $i$th phenotype individual, we allow the daughter cell to have the $j$th phenotype with a probability $A_{i,j}$.
}
Following nutrient depletion, the system transitions to a famine period, during which populations decline.
We denote the growth rate and death rate of phenotype $i$ in environment $e$ by $\mu_{i, e}$ and $\gamma_{i, e}$, respectively. 
For a single environmental period, the model equation for the population of phenotype $i$, $X_i$, is as follows: 
\begin{eqnarray}
  \frac{dX_i}{dt} = 
  \begin{cases}
      \left(1-\sum\limits_{i \neq j}{A_{i,j}}\right)\mu_{i,e}X_{i}
      + \sum\limits_{i \neq j}{A_{j,i}\mu_{j,e}X_{j}} & \text{($S>0$)},\\[1.4em]
    -\gamma_{i,e}X_{i} & \text{($S=0$)},
  \end{cases}
  \label{N_growth}
\end{eqnarray}
\begin{eqnarray}
  \frac{dS}{dt} = \begin{cases}
    -\sum_{i=1}^{N}{\mu_{i,e}X_{i}}& \text{($S>0$)},\\
    0 & \text{($S=0$)},\\
  \end{cases}
  \label{N_nutrient}
\end{eqnarray}
where $S$ is the amount of nutrient. For simplicity, we assumed that the amount of nutrients was set to the type-independent value $S_0$ at the supply event.  
As noted earlier, the supplied nutrient is almost always exhausted well before the next pulse; accordingly, 
at most one nutrient type is present at any time, and no leftovers are carried over in our setup.
As the type of supplied nutrient is encoded in the growth and death rate, we omit the subscript on $S$ to denote the type of nutrient and only represent the amount.
The following results are not affected if we generalize the setup such that the nutrients can be left unconsumed.
Because cells grow whenever nutrients remain, the population must experience the famine period for the population to reach a steady time-averaged level over a single feast-famine period. Thus, the supplied nutrients are almost always depleted before the next nutrient supply event after the population dynamics reach a steady state at the time-average level.

During the feast ($S>0$) period, cells grow and proliferate at a rate $\mu_{i,e}$. During the famine ($S=0$) period, cells die at a rate $\gamma_{i,e}$.
While we adopted a differential equation framework to describe continuous-time population dynamics, we incorporated the discreteness of the population by applying a threshold $\theta$ at each nutrient supply event. Any phenotype with a population size below threshold $\theta$ was set to zero.
This threshold was evaluated only at nutrient supply events.
Note that $X_i=0$ is not the absorbing state for the $i$th phenotype because the population is supplied from other phenotypes by the phenotypic switching process.

Nutrient supply events occur at random and discrete time points $\tau_k$ ($k=1,2,\ldots$). 
At each event, the amount of nutrient $S$ discontinuously jumps to a constant value $S_0$. 
\begin{eqnarray}
  &S(\tau_k) = S_0.
  \label{eq:NutrientSupply}
\end{eqnarray}
The type of supplied nutrient, that is, the environmental index $e$, was randomly chosen for the nutrient supply event $\tau_k$ (Fig~\ref{fig:example1}A).

In this model, we introduced two types of trade-offs. The first trade-off is the resource-use trade-off. The resource-use trade-off is often classified into the strong- and weak ones based on the functional form of the trade-off. In the present paper, we study the population dynamics with a strong trade-off. In particular, the trade-off strength is characterized by the curvature of a trade-off function between two traits, $y=f(x)$, where $x$ and $y$ represent cellular traits that cannot be simultaneously optimized: convex functions correspond to strong trade-offs, whereas concave functions correspond to weak trade-offs \cite{egasEvolutionRestrictsCoexistence2004,ehrlichTraitFitnessRelationships2017}. 

\red{
According to a prior work by Caetano and colleagues \cite{caetanoEvolutionDiversityMetabolic2021}, weak trade-offs inherently favor generalists and leave no room for strategy diversification. In our framework, introducing a growth–death trade-off would only reinforce this generalist dominance, because fast-growing specialists would suffer a larger death rate during famine. Thus, to establish a theoretical setting where different strategies can potentially compete, we introduced a strong trade-off between the growth rates on different nutrients using the following function:
}
\begin{eqnarray}
  \left(\prod_{e=1}^{E}\mu_{i,e}\right)^{1/E} = \bar{\mu}.\label{eq:resrTradeoff}
\end{eqnarray}

In addition to the above trade-off (Eq~(\ref{eq:resrTradeoff})), we introduce a trade-off between growth and death rates.
In a study by Biselli et al. \cite{biselliSlowerGrowthEscherichia2020},
the authors cultured {\it{E. coli}} cells in several flasks containing distinct sole carbon sources, and then the cells were washed and transferred into a carbon-free buffer medium to examine the relationship between growth rate under nutrient-rich conditions
and death rate during the subsequent starvation state.
They reported that the death rate in a carbon-free buffer medium 
increases exponentially with the growth rate under nutrient-rich preculture conditions.
This relationship was generic for the different carbon sources in the medium.
According to this result, we introduce an exponential trade-off between the growth rate and the death rate as follows:
\begin{eqnarray}
  \gamma_{i,e} = a \cdot \exp(b\mu_{i,e}),\label{eq:growth-death}
\end{eqnarray}
where $a$ and $b$ are constant values independent of the indices of phenotype ($i$) and environment ($e$), respectively.

\red{
Note that the main results being presented in the following sections are not qualitatively altered if we utilize different functional forms for the trade-offs. We confirmed that using alternative functions yields consistent results, as long as the resource-use trade-off (Eq~(\ref{eq:resrTradeoff})) remains convex and the growth-death trade-off (Eq~(\ref{eq:growth-death})) remains supralinear (see \nameref{sec:SI} Sections 1 and 7, respectively). Furthermore, we also verified that the main result is qualitatively unchanged by the relaxation of assumptions: the same arguments hold even if we adopt Monod-type growth kinetics instead of the substrate concentration-independent growth rate, and also, the no-leftover assumption of the nutrients is not crucial for the main results (see \nameref{sec:SI} Sections 5 and 6 ).}

\section*{Results}
\subsection*{Transition between Generalist-dominance and Specialist-dominance}
The purpose of this study was to identify the factors that drive the diversification of nutrient utilization strategies in temporally varying environments. 
As an initial step for answering this question, we start with a simple setup, where the model consists of $N=3$ phenotypes and $E=2$ environmental conditions. We index the phenotypes by Arabic numbers (phenotypes 1, 2, and 3) and environments by alphabet (environments A and B). 

\begin{figure}[h!]
  \ifshowfig\includegraphics[]{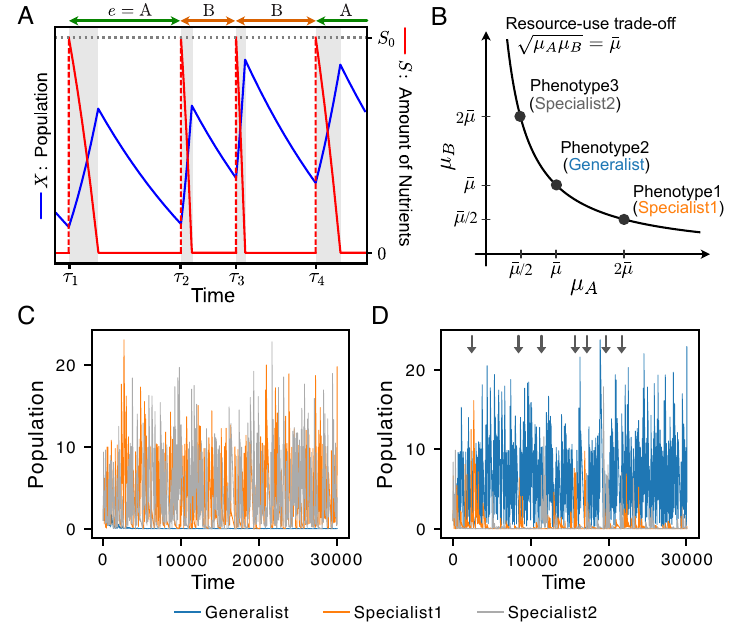}\fi
  \caption{\red{{\bf Schematic of the population dynamics model and examples of the simulation with $N=3$ and $E=2$.} } 
    \red{(A)  Temporal changes in population size and nutrient levels in a single phenotype ($N=1$) case. The blue line represents the population $X$, 
    and the red dashed line represents the amount of nutrients $S$. At the time points $\tau_1,\tau_2,\cdots$, 
    nutrients are supplied and $S$ is set to $S_0$.
    The feast phase (gray shaded area) begins at each nutrient supply event 
    and ends when the supplied nutrient is exhausted by cellular consumption, 
    after which the system enters the famine phase.
    At each time $\tau_k$, a different type of nutrient is randomly selected and supplied. 
    The type of nutrient supplied depends on the value of environmental variables.
    (B) The generalist (phenotype 2) is at $(\bar{\mu},\bar{\mu})$; the two specialists (phenotypes 1 and 3) are symmetric at $(2\bar{\mu},\bar{\mu}/2)$ and $(\bar{\mu}/2,2\bar{\mu})$, respectively.  
    (C) The case with $\bar{\mu}=0.4$. Two specialists dominate the population. (D) The case with $\bar{\mu}=0.8$. The generalist dominates the population. 
    The initial values of populations are set to unity for all three phenotypes.
    The parameters are set to $a=0.01,b=1,S_0=10,p=10^{-4}$, and $\theta=10^{-8}$.
    At each nutrient supply event, one of the two nutrient types (A or B) is chosen at random.
    The waiting time for the nutrient supply $\Delta\tau_k\;(=\tau_{k+1}-\tau_{k})$ follows a gamma distribution with shape and rate parameters being $2$ and $50$, $\Delta\tau_{k} \sim \Gamma(2,50)$.}
  }
  \label{fig:example1}
\end{figure}

In this simple setup, we set the growth rates of phenotypes in ascending and descending order, that is, the growth rate in environment A follows $\mu_{1,A}>\mu_{2,A}>\mu_{3,A}$ and that in environment B follows $\mu_{1,B}<\mu_{2,B}<\mu_{3,B}$. 
To begin with a simple case, we assume symmetry between environments A and B such that $\max_i{\mu_{i, A}}=\max_i{\mu_{i, B}}$ and $\min_i{\mu_{i, A}}=\min_i{\mu_{i, B}}$. 
One way to implement this symmetry consistent with the resource-use trade-off (Eq~(\ref{eq:resrTradeoff})) is as follows:
\[	
\mu_{1, A}=2\bar{\mu},\quad \mu_{1,B}=\bar{\mu}/2,\quad 
\mu_{2, A}=\mu_{2, B}=\bar{\mu},\quad
\mu_{3, A}=\bar{\mu}/2,\quad\mu_{3, B}=2\bar{\mu}.
\]
Because of this growth rate setting, phenotype 2 has a generalist-like strategy, where the growth rates are the same in the two environments, whereas the other phenotypes are specialist-like compared to phenotype 2 (see Fig~\ref{fig:example1}B). 
Note that the two specialists can also grow under environmental conditions in which the phenotypes have a lower growth rate, therefore, they are not extreme specialists who cannot grow under certain environmental conditions.

We assumed that the choice of the nutrient utilization strategy did not affect the probability of phenotypic switching. In addition, the phenotypic switch is considered to occur only in the nearest neighbor of the phenotype index. Overall, the phenotype switching probabilities are given by
\begin{equation}
A_{ij}=
\begin{cases}
p&(|i-j|=1),\\
0&(\text{otherwise}).
\end{cases}
\end{equation}

We set the waiting time for nutrient supply events $\Delta\tau_{k}$ to follow a gamma distribution.
The gamma distribution is a generalized form of the Erlang distribution that is widely employed to describe the waiting-time distribution of events that follow the Poisson process. 
\red{At each nutrient supply event, one of the two nutrient types (A or B) is chosen at random.}

Figs~\ref{fig:example1}C and D show the population dynamics in two different simulation setups, the geometric mean of the growth rates, $\bar{\mu}=0.4$ and $0.8$.
We observed two distinct types of population dynamics: in the first case, two specialist phenotypes dominated the community (Fig~\ref{fig:example1}C), and in the second case, the generalist phenotype became dominant (Fig~\ref{fig:example1}D). 
\red{The arrows in Fig~\ref{fig:example1}D indicate a transient increase in the specialist population, which will be discussed later.}

\begin{figure}[h!]
  \ifshowfig\includegraphics[]{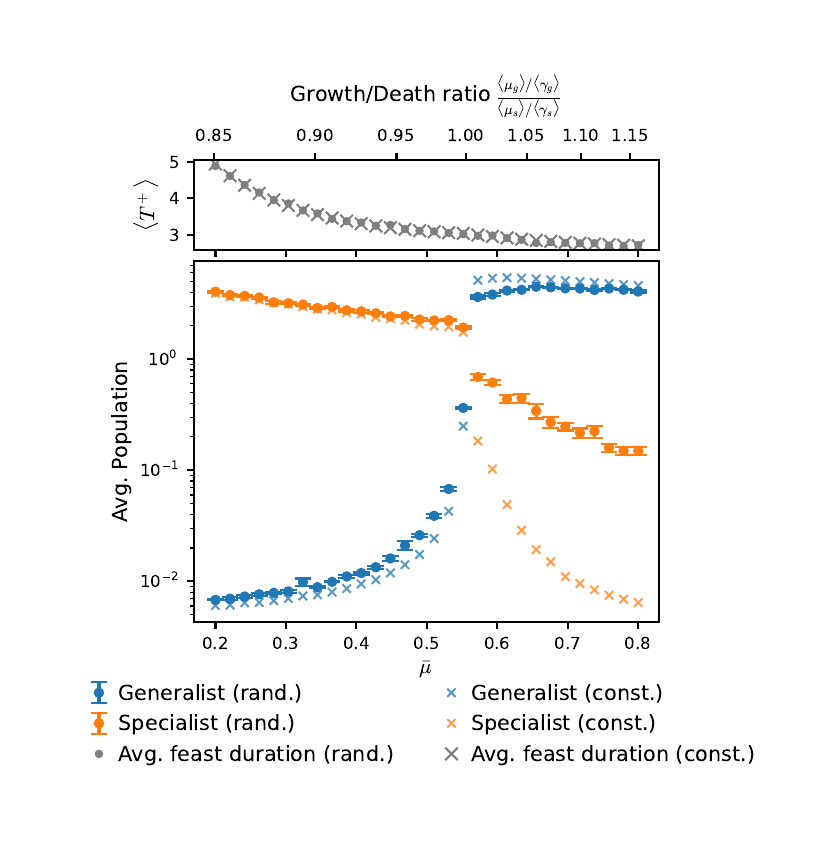}\fi
  \caption{
    \red{{\bf Transition between generalist and specialist dominance.}  
    Temporally averaged populations of the one generalist and two specialists as a function of $\bar{\mu}$. Crosses (\scalebox{0.8}{$\times$}) denote results with constant nutrient supply intervals ($\Delta\tau_{\rm{const.}}=100$), whereas circles (\textbullet) denote those with gamma-distributed intervals ($\Delta\tau\sim\Gamma(2,50)$). \red{Above this, the adjacent top panel illustrates the mean duration of the feast phase, $\langle T^{+} \rangle$, for the respective conditions.} The secondary top $x$-axis displays the ratio $(\langle\mu_g \rangle/\langle\gamma_g \rangle)/(\langle\mu_s \rangle/\langle\gamma_s \rangle)$. 
    Here, $\langle\mu_g \rangle$ ($\langle\mu_s \rangle$) and $\langle\gamma_g \rangle$ ($\langle\gamma_s \rangle$) denote the arithmetic means of the growth and death rates of the generalist (specialist) across the two environmental conditions. 
    Populations were averaged from $t=2.5\times10^5$  to $t=3.0\times10^5$ across 10 simulations; error bars indicate standard error. 
    At each nutrient supply event, one of the two nutrient types (A or B) is chosen at random. All other parameters follow Fig~\ref{fig:example1}C and D.}
    }
  \label{fig:avePlot}
\end{figure}

To study how the dominant phenotypes changed depending on the geometric mean of the growth rate $\bar{\mu}$, we simulated the model with various values of $\bar{\mu}$. 
Circle marks in Fig~\ref{fig:avePlot} show the temporal average of the population of the generalist and two specialists when $\Delta\tau_k$ follows a gamma distribution. The temporal average of the two specialists was taken, and the mean of the two averages is plotted in the figure.
The generalist is dominant for a larger $\bar{\mu}$, and the specialists are dominant for a smaller $\bar{\mu}$. 
\red{We also performed simulations with constant $\Delta\tau_k$. Cross marks indicate the corresponding temporally averaged population; representative time series with constant $\Delta\tau_k$ are shown in \nameref{fig:S1fig}.
We will revisit the gap between the specialist populations under random $\Delta\tau_k$ and constant $\Delta\tau_k$ later in the paper.}

To elucidate the mechanism of the transition between generalist and specialist dominance, we introduce a simple dimensionless parameter given by
$r=\langle\mu\rangle/\langle\gamma\rangle$, where $\langle\mu\rangle $ and $\langle\gamma\rangle$ are given by 
$\langle\mu\rangle = (\mu_A + \mu_B )/2$ and $\langle\gamma\rangle=(\gamma_A+\gamma_B)/2$, respectively.
Because the two specialist phenotypes exhibited symmetric growth properties with respect to the two environments ($\mu_{1, A}=\mu_{3, B}$ and $\mu_{1, B}=\mu_{3, A}$), the values of $r$ for the two specialists were the same.
We denote the $r$ values of the specialists and generalists as $r_s$ and $r_g$, respectively. The ratio of $r_g$ to $r_s$ was used as the horizontal top axis in Fig~\ref{fig:avePlot}. 
The figure shows that the transition between generalist- and specialist-dominant dynamics occurs at the value of $r_g/r_s$ being unity.
The value of $r_g/r_s$ at the transition point is robust to the parameter values of the phenotype switching probability $p$ (\nameref{fig:S2fig}). 
The transition of dominant phenotypes was well captured by the ratio of the $r$-value between generalists and specialists.

We carried out the same analysis while varying $b$, which controls the strength of the growth-death trade-off (\nameref{fig:S4fig}~Panel A).
\red{We observed a transition between specialist-dominant and generalist-dominant dynamics, also while changing the trade-off strength of the growth rate and death rate $b$.}

\red{
To understand why the ratio $r_g / r_s$ predicts the dominant phenotypes, we analyze a simplified version of the original model (Eqs~\eqref{N_growth} and \eqref{N_nutrient}). We apply a coarse-graining approach to the two environmental conditions, considering a single ``averaged'' environment. In this environment, the growth and death rates of each phenotype are defined as the arithmetic means of those in the two original environments. Specifically, for phenotype $i$, the average growth rate $\mu_i$ and death rate $\gamma_i$ are given by:}
\begin{eqnarray}
\mu_{i} = \frac{\mu_{i,A} + \mu_{i,B}}{2}, 
\gamma_{i} = \frac{\gamma_{i,A} + \gamma _{i,B}}{2}.
\end{eqnarray}
\red{
Since the two specialists in the original model are symmetric to the environments, after the coarse-graining of the environments, the two specialists are identical in terms of the growth rate and death rate. Thus, we deal only with the two phenotypes $\alpha$ and $\beta$.
}

\red{
To determine the capacity of each phenotype to invade under time-varying resource availability,
we quantified the overall population change across a single feast-famine cycle as the net logarithmic growth between two consecutive nutrient supply events ($\ln{(X_i(\tau_{k+1})/X_i(\tau_{k}))}$).
The net logarithmic growth of phenotype $i$, $f_i$ is given by
\begin{eqnarray}
f_{i} = \mu_{i}T^{+}(\bm X)-\gamma_{i}(\Delta\tau- T^{+}),\label{eq:f}
\end{eqnarray}
}
where $T^{+}$ is the duration of nutrient availability after the nutrient supply. If $f_i$ is positive, the population size increases, and if $f_i$ is negative, the population size decreases. \red{Note that the feast duration is set by how fast the individuals consume the supplied nutrient, $T^+$ is a function of $\bm X=(X_\alpha,X_\beta)$. }

\red{
Here, we study the invasibility of a phenotype $\beta$ to the steady population dominated by the phenotype $\alpha$. Suppose that the population dynamics is in the steady-state with only the phenotype $\alpha$. Thus, $f_\alpha=0$ holds, and from Eq~\eqref{eq:f} $T^+$ is calculated as
\begin{eqnarray}
T^{+}((X_\alpha,0)) = \frac{1}{\mu_\alpha}\ln\frac{S_0+X_\alpha}{X_\alpha}=\frac{1/\mu_{\alpha}}{1/\gamma_{\alpha}+1/\mu_{\alpha}}\Delta\tau.\label{eq:Tp}
\end{eqnarray}
We assume that during the attempt of invasion, the population of $\beta$, $X_\beta$ is sufficiently smaller than $X_\alpha$, and thus, $T^{+}((X_\alpha,X_\beta))\approx T^{+}((X_\alpha,0))$ holds. If the net logarithmic growth of the phenotype $\beta$, $f_\beta$, is positive with this $T^{+}(\bm X)$, the phenotype $\beta$ can invade into the population. The calculation is straightforward and $f_\beta$ is given by
\begin{eqnarray}
f_{\beta} = 
\gamma_\beta\frac{1/\mu_\alpha}{1/\mu_{\alpha}+1/\gamma_{\alpha}}\Delta\tau\left(\frac{\mu_\beta}{\gamma_\beta}-\frac{\mu_\alpha}{\gamma_\alpha}\right).
\end{eqnarray}
}
Note that the prefactor before the parenthesis is always positive, and thus, phenotype $\beta$ can invade if $\mu_{\beta}/\gamma_{\beta }>\mu_{\alpha }/\gamma_{\alpha }$ holds.

\red{
These invasion calculations provide a mechanistic explanation for the simulation results in Fig~\ref{fig:avePlot}.
In particular, a phenotype can invade a resident population when its growth-death rate ratio is larger than that of the resident phenotype (see \nameref{sec:SI} Section 2 for details).
As the invasive phenotype increases, nutrients are consumed more rapidly, which reduces $T^{+}$ and lowers the resident's net logarithmic growth.
The invasion proceeds until $T^{+}$ decreases to the point where the invasive phenotype satisfies $f_{\beta}=0$; at this point, the resident phenotype has negative $f_{\alpha}$ and its population declines.
Therefore, the transition between generalist and specialist dominance occurs at $r_g/r_s=1$.
}

Interestingly, the transition did not occur in a simulation with one type of specialist and one type of generalist (i.e., phenotype 3 excluded) (\nameref{fig:S3fig} and \nameref{fig:S4fig}~Panel B). 
In such cases, a generalist always dominates the system, and thus, the transition from generalist dominance to specialist dominance takes place through the cooperation of specialists to suppress the increase in the generalist population.  

\subsection*{Specialists have an advantage under sparse and uncertain nutrient supply}
\subsubsection*{Mean of nutrient supply interval}
So far, we have focused on the effect of $\bar{\mu}$ on population dynamics, which determines the growth-death rate ratios of the generalist and the specialists, $r_g$ and $r_s$. \red{However, Fig~\ref{fig:avePlot} suggests that under generalist-dominant conditions, the temporally averaged population of specialists increases when $\Delta\tau$ is random compared to when it is constant. Motivated by this observation,} we investigate the impact of $\Delta\tau$ on population dynamics, starting with the effect of the mean $\Delta\tau$.

\begin{figure}[h!]
  \ifshowfig\includegraphics[]{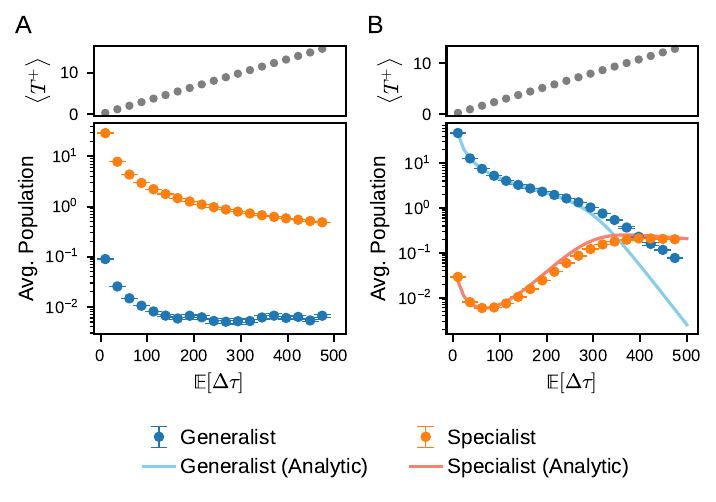}\fi
  \caption{\red{{\bf Temporal average of the population as a function of $\mathbb{E}[\Delta\tau]$.} 
  Changes in the temporally averaged populations as $\mathbb{E}[\Delta\tau]$ is varied in simulation with one generalist and two specialists. 
  The case with $\bar{\mu}=0.4$ is shown in (A) and the case with $\bar{\mu}=0.8$ is shown in (B).
  Above these, the adjacent top panels illustrate the mean duration of the feast phase, $\langle T^{+}\rangle$, for the respective conditions. 
  At each nutrient supply event, one of the two nutrient types (A or B) is chosen at random.
  We fixed the variance ($\rm{Var}(\Delta\tau) = \mathbb{E}[(\Delta\tau-\mathbb{E}[\Delta\tau])^2]=50$) and used different distributions with varying means.
  For each parameter, we ran 10 simulations and averaged the results.
  The error bars indicate the standard error. All other parameters were identical to those used in Fig~\ref{fig:example1}. \red{The solid lines represent the approximated analytical solutions, with further details on their derivation provided in \nameref{sec:SI} Section 4.}}
  }
  \label{fig:avePlotDeltaTau}
\end{figure}

Fig~\ref{fig:avePlotDeltaTau} shows the temporal average of the population of the generalist and two specialists as a function of the mean of nutrient supply interval, \red{$\mathbb{E}[\Delta\tau]$. In these simulations, we sampled $\Delta\tau$ from probability distributions with different means while keeping their variance fixed.}

The effect of $\mathbb{E}[\Delta\tau]$ on the temporally averaged population depends on whether the system is in the specialist-dominant or generalist-dominant condition, 
which is determined by the value of $\bar{\mu}$.
In the case of $\bar{\mu}=0.4$ (Fig~\ref{fig:avePlotDeltaTau}A, specialist-dominant condition), 
the temporally averaged population of the generalist and the specialists declined as $\mathbb{E}[\Delta\tau]$ increased.
This decline is expected because each supply event delivers a fixed amount of nutrient $S_0$, so the time-averaged nutrient supply rate decreases as $S_0/\mathbb{E}[\Delta\tau]$.

On the other hand, in the case of $\bar{\mu}=0.8$ (Fig~\ref{fig:avePlotDeltaTau}B, generalist-dominant condition),
\red{
the temporally averaged population of the specialists increased within a certain range of $\mathbb{E}[\Delta\tau]$ ($100\lessapprox \mathbb{E}[\Delta\tau]\lessapprox 300$),
while that of the generalist continued to decline. 
For $\mathbb{E}[\Delta\tau]$ being larger than $\approx 400$, the temporally averaged population of the specialists exceeded that of the generalist. 
The solid lines in Fig~\ref{fig:avePlotDeltaTau}B represent an analytical estimate of the populations under the generalist-dominating and constant $\Delta\tau$ conditions. The detailed derivations and results are provided in \nameref{sec:SI} Section 4. 
}

\red{
In the large-$\mathbb{E}[\Delta\tau]$ regime, the analytical solution for the generalist deviates substantially from the simulation results.
This discrepancy arises from an assumption used in the analytical calculation, namely that the environmental condition alternates deterministically between the two states. Indeed, when we run simulations under deterministic alternation, the results agree closely with the analytical solution (see \nameref{sec:SI} 4 for details).
}




\red{
The specialist population exhibits a characteristic behavior where its size increases with the expected interval between nutrient supply events, $\mathbb{E}[\Delta \tau]$. This dependence is counterintuitive, as an increase in the starvation interval would typically be expected to exert only negative impacts on the populations.
}

\red{
However, this phenomenon can be understood by considering how longer intervals allow specialists to consume supplied nutrients more effectively than generalists. Following a prolonged famine period, the populations of all phenotypes—whether generalist or specialist—decline significantly. The critical point lies in the state of the system after this strong decline: at the subsequent nutrient supply event, the population of every phenotype is sufficiently small that the initial abundance of the previously dominant phenotype no longer provides a competitive advantage.
}

\red{
Under these conditions, the growth rate $\mu_{i,e}$ becomes the primary determinant of nutrient acquisition. In each environmental condition, one of the two specialists possesses the highest growth rate among the three phenotypes. While the generalist dominates the system most of the time, a longer interval reduces the absolute difference in population sizes between the generalist and the specialists. Consequently, specialists can utilize nutrients that would otherwise have been consumed by the generalists, leading to an increase in the specialist population as the interval $\mathbb{E}[\Delta \tau]$ grows. 
}

\red{
In the regime $100 \lessapprox \mathbb{E}[\Delta \tau] \lessapprox 300$ shown in Fig~\ref{fig:avePlotDeltaTau}B, the observed increase in the specialist population can be analytically approximated as:
\begin{equation}
X^{\rm approx.}_s(\Delta\tau_{\rm const.}) \sim \frac{\exp(d \Delta \tau_{\rm const.})}{\Delta \tau_{\rm const.}}, \label{eq:exp}
\end{equation}
where $d$ is a positive constant. In this estimation, we employ $\Delta \tau_{\rm const.}$ instead of $\mathbb{E}[\Delta \tau]$ by assuming a constant interval for analytical tractability (see \nameref{sec:SI} Section 4 for the details of calculation).
}

\red{
Eq \eqref{eq:exp} encapsulates two competing effects. The factor $\Delta \tau_{\rm const.}^{-1}$ represents the decrease in population density resulting from the reduced frequency of nutrient supply per unit time. Conversely, the exponential term arises from the condition where a specialist consumes nearly all available nutrients during a single supply event. Specifically, the population growth during a feast period is given by $\exp(\mu_{i,e} T^+)$. Since the duration of the feast period $T^+$ scales linearly with the interval $\Delta \tau$ (refer to Eq~\eqref{eq:Tp}), the specialist's growth within a single cycle exhibits an exponential dependence on $\Delta \tau$.
}

\red{
However, as $\Delta\tau_{\rm{const.}}$ increases further ($\mathbb{E}[\Delta\tau] \gtrapprox 350$ in Fig~\ref{fig:avePlotDeltaTau}B), the difference in population size between the generalist and the specialists becomes negligible at the onset of every nutrient supply event. This convergence of population sizes effectively terminates the growth mechanism characteristic of the intermediate $\mathbb{E}[\Delta\tau]$ regime. In that intermediate regime, the specialists increase their population by sequestering nutrients that the generalist would otherwise consume.
}

\red{
In the large $\mathbb{E}[\Delta\tau]$ region, however, this capacity for specialists to further displace the generalist's nutrient consumption is exhausted. Because the specialists can no longer expand their share of nutrient acquisition at the expense of the generalist, the population dynamics revert to the $\mathbb{E}[\Delta\tau]$ trend, primarily driven by the overall reduction in the frequency of resource supply.
}


\red{
Note that the inversion at $\mathbb{E}[\Delta\tau] \approx 400$ in Fig~\ref{fig:avePlotDeltaTau}B does not contradict the $r_g/r_s$ criterion. This crossover arises because time-averaging captures the specialists' transient, high-amplitude growth bursts during nutrient pulses, which inflate their mean population despite sharp declines during famines. The result reflects these short-lived peaks rather than a reversal of the fundamental evolutionary dominance determined by $r$.
}


\subsubsection*{Variance of nutrient supply interval}
\red{
We have so far examined how the mean nutrient supply interval affects time-averaged population sizes. However, Fig~\ref{fig:avePlot} reveals that these averages differ between constant and random interval simulations, even when the mean interval remains identical. Specifically, under generalist-dominant conditions, specialists exhibit higher temporal averages when $\Delta\tau$ follows a Gamma distribution (circles) compared to the constant-interval case (crosses). By contrast, in specialist-dominant conditions, the corresponding increase in the generalist population is smaller.
Indeed, the time courses in Fig~\ref{fig:example1} support this interpretation.
Generalists rarely increased in the specialist-dominant condition (Fig~\ref{fig:example1}C).
On the other hand, specialists often showed transient increases in the generalist-dominant condition (arrows in Fig~\ref{fig:example1}D).
}

\red{
The analytical estimate also provides insights into the effect of variance in $\Delta\tau$.
As described in the previous subsection, our analytical solutions for the constant-interval case show that the specialist population has the form $\exp(d\Delta\tau_{\rm{const.}})/\Delta\tau_{\rm const.}$ (Eq ~\eqref{eq:exp}) within a range where the specialist population increases with the mean interval. Although the expression is obtained with the constant interval, we heuristically assume that the population size of the specialist quickly relaxes to the expression while $\Delta\tau$ is randomly changing and calculate the population of the specialist averaged over the distribution of $\Delta\tau$ as 
\begin{equation}
\langle X^{\rm approx.}_s(\Delta \tau)\rangle =\lim_{n\to\infty}\frac{\sum_{k=1}^n X^{\rm approx.}_s(\Delta \tau_k)\Delta \tau_k}{\sum_{k=1}^n \Delta \tau_k}=\frac{\mathbb E[X^{\rm approx.}_s(\Delta \tau)\Delta\tau]}{\mathbb E[\Delta\tau]},
\end{equation}
where $\Delta \tau_k$ is the $k$th realization of the interval. Note that $\langle \cdot \rangle$ represents the temporal average of the population, but it is not the mean of $X^{\rm approx.}_s(\Delta \tau)$ over $\Delta\tau$ distribution. By performing the cumulant series expansion of the numerator, we have 
\begin{equation}
\langle X^{\rm approx.}_s(\Delta \tau)\rangle\sim\frac{\exp{\left(\sum_{n=1}^{\infty}\frac{d^{n}}{n!}\kappa_n\right)}}{\mathbb E[\Delta \tau]},
\end{equation}
where $\kappa_n$ is the $n$th cumulant of $\Delta\tau$. In the expansion, all expansion coefficients are positive and $\kappa_2$ corresponds to the variance. Therefore, in the regime where Eq~\eqref{eq:exp} holds, an increase in the variance of $\Delta\tau$ also enhances the temporally averaged population of specialists.}



\subsection*{Diversification of strategies in continuous strategy space and multiple environmental conditions}
\red{
In the previous sections, we have addressed the transition of the strategies between the generalist-type and the specialist-type. In the model (Eq~\eqref{N_growth}), we pre-define the difference of the growth rate, and accordingly, death rate among phenotypes. This setting, albeit simplifying the model, artificially restricts the population dynamics by limiting the possible strategies. Thus, in this section, we extend the model so that the individuals can change the strategies rather continuously. Finally, we study the model for more than two environmental conditions with continuous phenotype setting.
}

\red{
To make the phenotype space finer, we have increased the number of phenotypes ($N = 51$). As in the simulations with one generalist and two specialists, we set the growth rates of phenotypes in ascending and descending order. At this stage, we keep using the two-environment setup with symmetry between environments A and B.}
We set the growth rate of phenotype $i$ as follows, which is consistent with the resource-use trade-off.
\begin{eqnarray}
\mu_{i,A} = \bar{\mu}\cdot g_{i},\;\mu_{i,B} = \bar{\mu}/ g_{i},\; g_{i} = g^{(0)}\cdot m^{c\cdot i}.
\label{eq:growth2continuous}
\end{eqnarray}
Here, $\Delta\tau$ is set to a constant value, and the environment is changed alternately.

\begin{figure}[h!]
  \ifshowfig\includegraphics[]{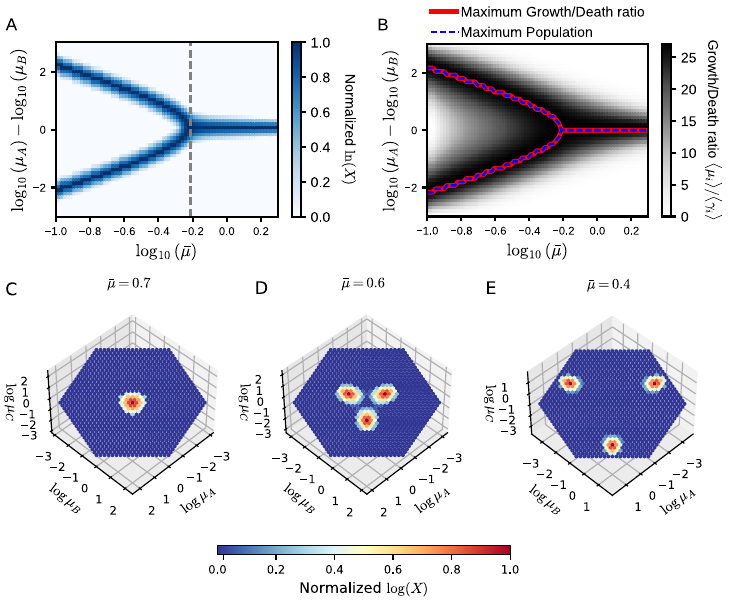}\fi
  \caption{\red{{\bf Normalized population density distribution.}  
(A) Simulation with two environmental conditions.
  The $x$-axis represents the geometric mean of growth rates, and the $y$-axis represents the difference in logarithmic growth rates between the two environments. We varied  $\bar{\mu}$ from $10^{-1.0}$ to $10^{0.3}$ and ran simulations until $t=2.0\times10^5$. Population sizes were time-averaged over the last 20 nutrient supply intervals. Growth rates are defined by Eq~(\ref{eq:growth2continuous}) with $m=2$, $c=0.2$ and $g^{(0)}=m^{(-c(N+1)/2)}$.
  (B)  Plot of the growth-to-death ratio for each phenotype. The phenotypes with the highest growth-to-death ratio and with the largest population at a given $\bar{\mu}$ are highlighted by the solid red line and the dashed blue line, respectively. (C-E) Simulation with three environmental conditions.
  Each dot represents one phenotype.
  We simulated the population dynamics up to $t=2.0\times 10^{5}$.
  Population sizes were time-averaged over the last 30 nutrient supply intervals.
  The growth rates of each phenotype across the three environments, denoted as $(\mu_A,\mu_B,\mu_C)$, were set to $(\bar{\mu}\cdot 2^x,\bar{\mu}\cdot 2^y,\bar{\mu}\cdot 2^z)$, where $x+y+z=0$ holds. The phenotypes are discretely distributed on this plane such that adjacent phenotypes are generated by holding one parameter constant while increasing one of the remaining two by a fixed step size and decreasing the other by the same amount.
  Through panels A and C to E, the supplied nutrient type is switched deterministically, cycling through the specified nutrient set.
  For each simulation, the logarithm of the temporally averaged population in each bin, normalized from 0 to 1, is plotted.
  The parameters are set to $a = 0.01, b = 1, S_0 = 10$ and $\Delta\tau_{\rm{const.}} = 10$.
  Phenotypes can switch only between adjacent bins, with a transition probability of $10^{-4}$.}}
  \label{fig:e3}
\end{figure}

We simulated the model dynamics for different values of $\bar{\mu}$, and plotted the temporal average of the population in Fig~\ref{fig:e3}A.
As a result, a continuous shift of the dominant strategy was observed by changing the value of $\bar{\mu}$.
When $\bar{\mu}$ becomes larger than the branching point at $\bar{\mu}\approx 0.6$ (gray dashed line in Fig~\ref{fig:e3}A), the generalist-like strategy becomes dominant.
When the value of $\bar{\mu}$ falls below the branching point, the population abundance distribution in the phenotype space becomes bimodal.
The phenotypes at the local maxima of the distribution are symmetric between different environments, i.e., the growth rates of the two phenotypes, $\alpha$ and $\beta$, have the following relationship: $\mu_{\alpha, A}=\mu_{\beta, B}$ and $\mu_{\alpha, B}=\mu_{\beta, A}$, where A and B are different environments.
We also observed a shift from the generalist-like strategy to the two symmetric specialist-like strategies with decreasing $b$. \red{Thus, the discontinuity presented in Fig~\ref{fig:avePlot} is due to the gap among the phenotypes in the model, and the transition itself is continuous.}

\red{
The shift of the dominant strategy in this continuous version of the model is consistently explained by the average growth-to-death ratio $r_i$. In Fig~\ref{fig:e3}B, we present the growth-to-death ratio of phenotypes as a heatmap. We highlighted the most abundant phenotype and the phenotype with the highest growth-to-death ratio by the dashed blue line and the solid red line, respectively. As shown in the figure, the two lines overlap, and thus, the most abundant phenotype is that whose growth-to-death ratio is the highest among the phenotypes.}

\red{We also performed simulations under conditions where a symmetric pair was missing (\nameref{fig:S7fig}). Consistent with the $N=2$ case (\nameref{fig:S3fig}), the generalist was always dominant. Specialists could emerge when $\bar{\mu}$ was small, but their abundance never exceeded that of the generalist.}

Next, we conducted simulations under three environmental conditions ($E=3$) to observe the dynamics of the system under more than two environmental conditions.
As in the setup above, we discretized the phenotype space ($({\mu_A\mu_B\mu_C})^{1/3}=\bar{\mu}$) using finely spaced bins.

\red{
Also in the three environment case, there is a critical $\bar{\mu}$ on which the dominant strategy branches from the single generalist strategy to the strategies with environment-dependent growth rates. In Fig~\ref{fig:e3}C-E, the temporal average of the population of each phenotype is shown in heatmaps. Recall that the phenotypes are characterized by the growth rates in the environments. Since we have three environments and one trade-off relationship among the growth rates, the degree of freedom of the growth rates is equal to two, and the population distribution is presented on the two-dimensional growth rate surface in the three-dimensional growth rate space. 
}

\red{
With the highest $\bar{\mu}$ among the three panels ($\bar{\mu}=0.7$ among $\bar{\mu}=0.7, 0.6$, and $0.4$) the population distribution concentrates at the center, meaning that the growth rates of the most abundant phenotype are identical among all three environments, i.e., the generalist type strategy (Fig~\ref{fig:e3}C). This peak splits into the three clusters as $\bar{\mu}$ decreases (Fig~\ref{fig:e3}D) and approaches the edge of the surface by decreasing the growth rate in one of the three environments (Fig~\ref{fig:e3}E). Therefore, we obtained the two-environment specialists in the simulation. 
}

\red{
We have not found any parameter set under which one-environment specialists stably coexist. In simulations with four environments, only three-environment specialists emerged. Although we have not analyzed cases with more than four environments due to computational cost, we hypothesize that only phenotypes close to the generalist (with similar growth rates across all environments) or near-generalist specialists that lack growth in only a single environment will emerge.
}
\section*{Discussion}

In the present study, we have developed a microbial population-dynamics model for a feast-famine cycle to study nutrient-utilization strategies in temporally varying environments. In the model, phenotypes differ in their growth rates on each nutrient type. These growth rates are constrained by a resource-use trade-off that represents metabolic limitations.
Nutrients are supplied to the ecosystem as discrete, stochastic events. Once supplied, they are rapidly consumed by microbes, leading to a nutrient-depleted famine period. The alternation of nutrient-rich and nutrient-depleted conditions leads to a feast-famine cycle.
Under this cycle, cells face a trade-off between the growth rate in feast and death rate in famine.

\red{
We explored the dynamics using a simple setup with three phenotypes and two types of nutrients. We systematically varied two parameters, the geometric mean of the growth rate $\bar{\mu}$ and the strength of the growth-death trade-off $b$.
Although the resource-use trade-off was fixed to the same convex form, the dominant strategy shifted between generalists and specialists depending on parameter values.
This contrasts with prior theoretical studies suggesting that strong trade-offs favor specialists\cite{egasEvolutionRestrictsCoexistence2004,ehrlichTraitFitnessRelationships2017,caetanoEvolutionDiversityMetabolic2021}. In our model, nutrient availability varies over time in the feast-famine cycle, where a growth-death trade-off played a central role in shaping the optimal nutrient utilization strategy. Specifically, the phenotype with the larger ratio of the arithmetic mean growth rate to the arithmetic mean death rate becomes dominant in the population.
}

\red{
The role of the growth-death trade-off is particularly significant given that the underlying resource-use trade-off is often evolutionarily rigid.
According to classical study by R. Levins
\cite{levinsEvolutionChangingEnvironments1968}, 
the curvature of a trade-off between distinct environmental conditions in general depends on the organism's adaptive capacity relative to the magnitude of environmental variation. The adaptive capacity of resource utilization is considered to be limited compared to other biological functions such as stress response and motility. A large portion of the adaptive capacity for resource use is set by metabolism. As represented by the infeasibility of the simultaneous use of glycolysis and gluconeogenesis, there are several constraints originating from the structure of the metabolic reaction networks and the thermodynamic nature of the biochemical reactions\cite{bar-evenThermodynamicConstraintsShape2012,schinkGlycolysisGluconeogenesisSpecialization2022}. Such fundamental constraints act as rigid barriers, rendering the trade-off curvature a fixed structural property rather than a flexible trait.
}

\red{
In contrast, the growth-death trade-off may exhibit greater evolutionary flexibility compared to the resource-use trade-off. For instance, experimental evidence indicates that the shape of the growth-death relationship differs significantly between wild-type strains and {\it rpoS}-deletion mutants\cite{biselliSlowerGrowthEscherichia2020}. Since {\it rpoS} mutations are commonly found in both laboratory and natural populations \cite{notley-mcrobbRpoSMutationsLoss2002}, this suggests that the strength of the growth-death trade-off (parameter $b$) can be tuned through mutations on a single regulatory gene.}

\red{
According to the analysis of the present model, the inherent flexibility of the growth-death trade-off may offer a potential framework for understanding how generalist and specialist strategies emerge or shift within natural populations. While our study relies on a simplified mathematical setup, these results suggest that such evolutionary tuning of growth-death trade-off could be one of the factors influencing the distribution of resource-utilization strategies in temporally varying environments.
}


We then investigated how the nutrient-supply interval $\Delta\tau$ affects population dynamics. Under generalist-dominant conditions, the temporally averaged population of specialists increased with both the mean and variance of the interval. This occurs because long intervals transiently depress total population size, extending the feast phase and thereby giving the faster-growing specialist a temporal advantage. \red{Even if the environmental condition favors generalist dominance, prolonged famine periods can significantly reduce the population density at the onset of the subsequent feast cycle. This transient relief from density-dependent competition allows specialists to exploit their higher growth rates more effectively before the population recovers to a level that exhausts the available nutrients. These results suggest that environmental intermittency can act as a potent driver for maintaining phenotypic diversity, preventing the complete exclusion of specialist traits in fluctuating ecosystems.}


Because the generalist phenotype, endowed with a high growth/death ratio, is evolutionarily stable, it persistently acts as a ``source,'' seeding rare specialist lineages that are unstable overall yet periodically advantageous (a ``sink''). Such source-sink dynamics have been discussed in both theoretical and empirical studies
\cite{
  laskaSinkHostAllows,
  perronSourceSinkDynamics2007,
  hermsenSourcesSinksStochastic2010,
  sriswasdiGeneralistSpeciesDrive2017
}.
These studies suggest that such dynamics can serve as key mechanisms for the long-term maintenance of diversity in ecological communities.

\red{
  This transient specialist advantage---which elevates the temporally averaged specialist abundance despite long-term generalist dominance---may also be analogous to the storage effect \cite{chessonMechanismsMaintenanceSpecies2000a}. In our context, phenotypic switching provides the ``storage'': it prevents specialist extinction during unfavorable periods, ensuring they remain present to exploit temporary opportunities created by reduced competition.
}

\red{
This study relates to recent mathematical modeling work on population dynamics in serial-dilution systems and environments with periodic nutrient supply\cite{manhartGrowthTradeoffsProduce2018, erezNutrientLevelsTradeoffs2020, fridmanFinescaleDiversityMicrobial2022, bloxhamDiauxicLagsExplain2022, bloxhamBiodiversityEnhancedSequential2024}.
Previous studies in this line have focused mainly on growth dynamics following nutrient addition (i.e., the ``feast'' in our framework). For example, these studies have explored how lag times, diauxic shifts, and concentration-dependent growth shape competition under repeated nutrient pulses. However, these frameworks do not explicitly treat population decline under nutrient depletion. In our framework, we extend this scope by coupling growth and death through a growth-death trade-off. Our results suggest that death during the famine phase plays an important role in determining the competitive outcome between different nutrient utilization strategies.
}

\red{
We simplified within-feast growth dynamics to isolate the effects of the trade-offs. Although our main findings remain robust to the detailed settings, several key factors remain to be explored in future work. Examples include patterns of multiple resource utilization (e.g., co-utilization or diauxic shifts) and utilization efficiency at low nutrient concentrations. 
Each of these factors can alter the dynamics of population and resources, promoting temporal niche differentiation that influences strategy selection and the maintenance of diversity in fluctuating environments\cite{manhartGrowthTradeoffsProduce2018,fridmanFinescaleDiversityMicrobial2022,bloxhamBiodiversityEnhancedSequential2024}. Incorporating these elements into our framework could provide alternative insights into the emergence of diverse nutrient utilization strategies. A systematic evaluation of how these factors operate in conjunction with growth-death and resource-use trade-offs to determine selection and diversification remains an important direction for future work.
}


\section*{Supporting information}
\paragraph*{S1 Fig.}\label{fig:S1fig}
\ifshowfig
\mbox{}\par
\begin{figure}[H]
\includegraphics[]{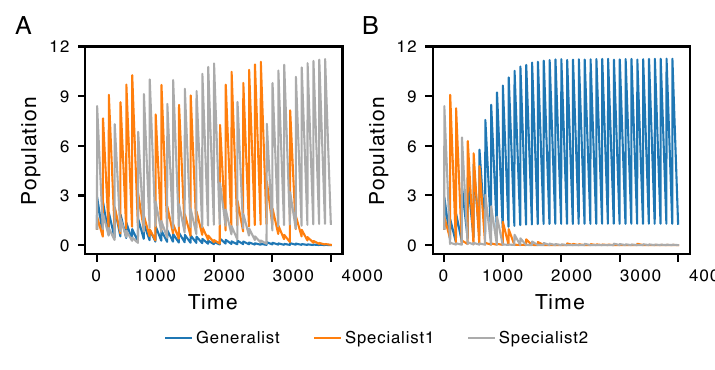}
\caption*{
  {\bf Time series plot of the population with constant $\Delta\tau$.}
  Examples of the simulation under the condition that nutrient supply events occur at regular intervals, $\Delta\tau_{\rm{const.}}=100$.
  (A) $\bar{\mu}=0.4$, (B) $\bar{\mu}=0.8$. 
  Nutrient supply events occur at regular intervals, $\Delta\tau_{\rm{const.}}=100$.
  At each nutrient supply event, one of the two nutrient types (A or B) is chosen at random.
  The other parameters were the same as those in Fig~\ref{fig:example1}.
}
\end{figure}
\else
\textbf{ Time series plot of the population with constant $\Delta\tau$.}
Examples of the simulation under the condition that nutrient supply events occur at regular intervals, $\Delta\tau_{\rm{const.}}=100$.(A) $\bar{\mu}=0.4$, (B) $\bar{\mu}=0.8$. Nutrient supply events occur at regular intervals, $\Delta\tau_{\rm{const.}}=100$.At each nutrient supply event, one of the two nutrient types (A or B) is chosen at random.The other parameters were the same as those in Fig~\ref{fig:example1}
\fi

\paragraph*{S2 Fig.}\label{fig:S2fig}
\ifshowfig
\mbox{}\par
\begin{figure}[H]
  \includegraphics[]{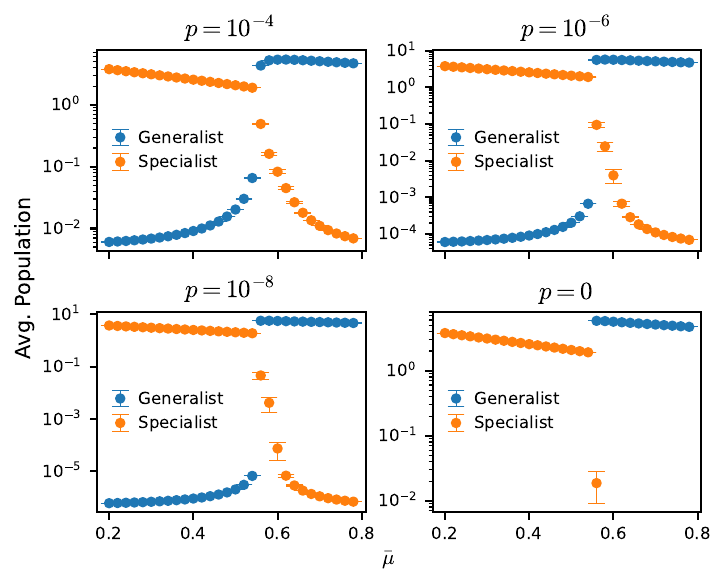}
\caption*{
  {\bf Effect of phenotype switching probability.}
Temporal average of the population with different phenotype switching probability between generalists and specialists, $p$.
Simulations were conducted with $p=10^{-4},10^{-6},10^{-8}$ and \red{$0$}. 
Nutrient supply events occur at regular intervals, $\Delta\tau_{\rm{const.}} = 100$.
At each nutrient supply event, one of the two nutrient types (A or B) is chosen at random.
The other parameter settings were the same as those in Fig~\ref{fig:avePlot}.
}
\end{figure}
\else
\textbf{ Effect of phenotype switching probability.}
Temporal average of the population with different phenotype switching probability between generalists and specialists, $p$.
Simulations were conducted with $p=10^{-4},10^{-6},10^{-8}$ and \red{$0$}. 
Nutrient supply events occur at regular intervals, $\Delta\tau_{\rm{const.}} = 100$.
At each nutrient supply event, one of the two nutrient types (A or B) is chosen at random.
The other parameter settings were the same as those in Fig~\ref{fig:avePlot}.
\fi
\paragraph*{S3 Fig.}\label{fig:S3fig}
\ifshowfig
\mbox{}\par
\begin{figure}[H]
\centering
\includegraphics[]{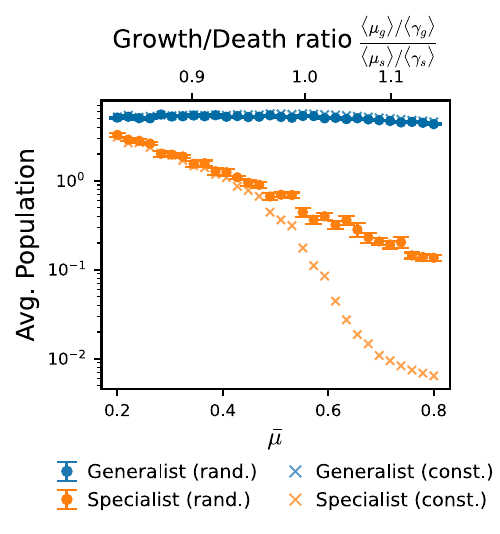}
\caption*{
  \red{{\bf Temporal average of the population with $N=2$.}
  Temporally averaged populations of the one generalist and one specialist as a function of $\bar{\mu}$. 
  Crosses (\scalebox{0.8}{$\times$}) denote results with constant nutrient supply intervals ($\Delta\tau_{\rm{const.}}=100$), 
  whereas circles (\textbullet) denote those with gamma-distributed intervals ($\Delta\tau \sim \Gamma(2,50)$).
  The secondary top $x$-axis displays the ratio $(\langle\mu_g \rangle / \langle\gamma_g \rangle)/(\langle\mu_s \rangle/\langle\gamma_s \rangle)$. 
  Here, $\langle\mu_g \rangle$ ($\langle\mu_s \rangle$) and $\langle\gamma_g \rangle$ ($\langle\gamma_s \rangle$) denote the arithmetic means of the growth and death rates of the generalist (specialist) across the two environmental conditions. 
  Populations were averaged from $t=2.5\times10^5$ to $t=3.0\times10^5$ across 10 simulations; error bars indicate standard error. 
  At each nutrient supply event, one of the two nutrient types (A or B) is chosen at random. All other parameters follow Fig~\ref{fig:avePlot}.}
}
\end{figure}
\else
\red{\textbf{Temporal average of the population with $N=2$.} Temporally averaged populations of the one generalist and one specialist as a function of $\bar{\mu}$. Crosses (\scalebox{0.8}{$\times$}) denote results with constant nutrient supply intervals ($\Delta\tau_{\rm{const.}}=100$), 
  whereas circles (\textbullet) denote those with gamma-distributed intervals ($\Delta\tau \sim \Gamma(2,50)$). The secondary top $x$-axis displays the ratio $(\langle\mu_g \rangle / \langle\gamma_g \rangle)/(\langle\mu_s \rangle/\langle\gamma_s \rangle)$. Here, $\langle\mu_g \rangle$ ($\langle\mu_s \rangle$) and $\langle\gamma_g \rangle$ ($\langle\gamma_s \rangle$) denote the arithmetic means of the growth and death rates of the generalist (specialist) across the two environmental conditions. 
  Populations were averaged from $t=2.5\times10^5$ to $t=3.0\times10^5$ across 10 simulations; error bars indicate standard error. At each nutrient supply event, one of the two nutrient types (A or B) is chosen at random. All other parameters follow Fig~\ref{fig:avePlot}.}
\fi
\paragraph*{S4 Fig.}\label{fig:S4fig}
\ifshowfig
\mbox{}\par
\begin{figure}[H]
\includegraphics[]{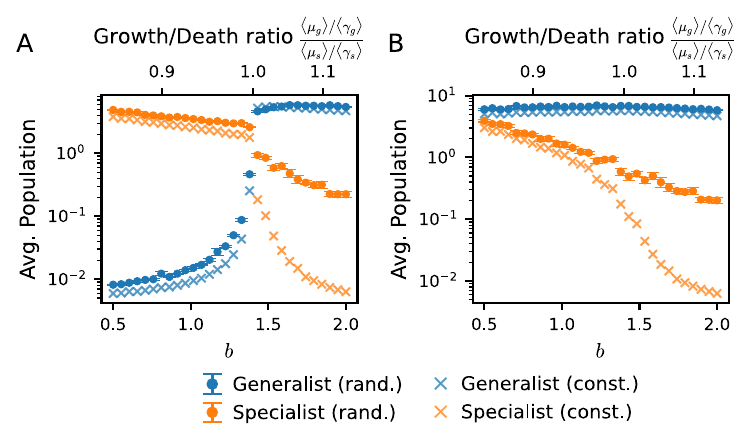}
\caption*{
  \red{{\bf Temporal average of the population.}
  The temporally averaged populations are shown as a function of the parameter $b$.
  The model with one generalist and two specialists is simulated for (A), while there is only a single specialist adapted to environment A for (B).
  The geometric mean of the growth rate $\bar{\mu}$ was set to 0.4.
  Crosses (\scalebox{0.8}{$\times$}) denote results with constant nutrient supply intervals ($\Delta\tau_{\rm{const.}}=100$), 
  whereas circles (\textbullet) denote those with gamma-distributed intervals ($\Delta\tau \sim \Gamma(2,50)$).
  The secondary top $x$-axis displays the ratio $(\langle\mu_g \rangle / \langle\gamma_g \rangle)/(\langle\mu_s \rangle/\langle\gamma_s \rangle)$. 
  Here, $\langle\mu_g \rangle$ ($\langle\mu_s \rangle$) and $\langle\gamma_g \rangle$ ($\langle\gamma_s \rangle$) denote the arithmetic means of the growth and death rates of the generalist (specialist) across the two environmental conditions. 
  At each nutrient supply event, one of the two nutrient types (A or B) is chosen at random.
We simulated up to $t = 3.0 \times 10^5$ and averaged the populations after $t = 2.5 \times 10^5$. 
For each parameter set, we ran 10 simulations and averaged the results. The error bars indicate the standard error.
The other settings were the same as those in Fig~\ref{fig:avePlot}.}
}
\end{figure}
\else
\red{\textbf{ Temporal average of the population.} The temporally averaged populations are shown as a function of the parameter $b$. The model with one generalist and two specialists is simulated for (A), while there is only a single specialist adapted to environment A for (B). The geometric mean of the growth rate $\bar{\mu}$ was set to 0.4. Crosses (\scalebox{0.8}{$\times$}) denote results with constant nutrient supply intervals ($\Delta\tau_{\rm{const.}}=100$), whereas circles (\textbullet) denote those with gamma-distributed intervals ($\Delta\tau \sim \Gamma(2,50)$). The secondary top $x$-axis displays the ratio $(\langle\mu_g \rangle / \langle\gamma_g \rangle)/(\langle\mu_s \rangle/\langle\gamma_s \rangle)$. Here, $\langle\mu_g \rangle$ ($\langle\mu_s \rangle$) and $\langle\gamma_g \rangle$ ($\langle\gamma_s \rangle$) denote the arithmetic means of the growth and death rates of the generalist (specialist) across the two environmental conditions. At each nutrient supply event, one of the two nutrient types (A or B) is chosen at random. We simulated up to $t = 3.0 \times 10^5$ and averaged the populations after $t = 2.5 \times 10^5$. For each parameter set, we ran 10 simulations and averaged the results. The error bars indicate the standard error. The other settings were the same as those in Fig~\ref{fig:avePlot}.}
\fi

\paragraph*{S5 Fig.}\label{fig:S5fig}
\ifshowfig
\mbox{}\par
\begin{figure}[H]
\includegraphics[]{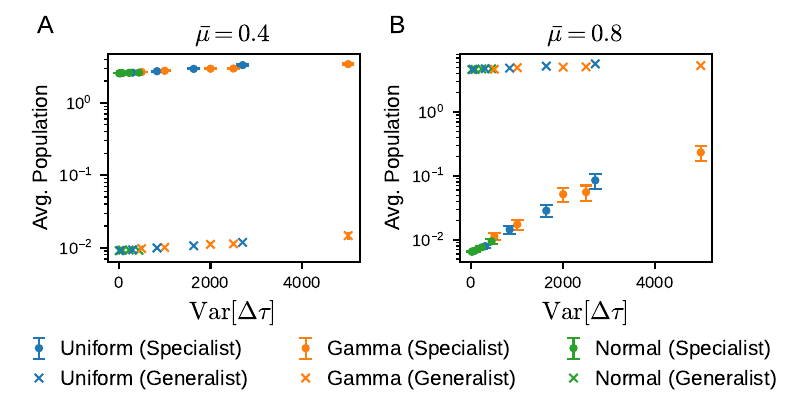}
\caption*{
  {\bf Effect of $\Delta\tau$ variance on the temporal average of the population.}
Temporal average of the population with different variances in $\Delta\tau$. 
We performed simulations with $\Delta\tau$ following uniform, normal and gamma distributions.
The dots represent the temporal average of the population of specialists, and the crosses represent those of generalists.
The geometric mean of the growth rate $\bar{\mu}$ differed between the two figures: $\bar{\mu}=0.4$ in (A) and $\bar{\mu}=0.8$ in (B).
For each of the three distribution families, 
we generated several variants with the same mean ($\mathbb{E}[\Delta\tau]=100$) but different variances and ran the simulations.
Uniform distributions were sampled from $U(100-d,100+d)$; for $d=10,30,50,70$ and $90$.
Normal distributions were sampled from $\mathcal{N}(100,\sigma^2)$; for $\sigma=3,9,15$ and $27$.
Gamma distributions were sampled from $\Gamma(k,100/k)$; for $k=2,4,5,10$ and $20$.
At each nutrient supply event, one of the two nutrient types (A or B) is chosen at random.
For each distribution, we performed 20 simulations and averaged the results.
The error bars indicate standard deviation.
All the other parameters were identical to those used in Fig~\ref{fig:avePlot}.
}
\end{figure}
\else
\textbf{Effect of $\Delta\tau$ variance on the temporal average of the population.}
Temporal average of the population with different variances in $\Delta\tau$. 
We performed simulations with $\Delta\tau$ following uniform, normal and gamma distributions.
The dots represent the temporal average of the population of specialists, and the crosses represent those of generalists. The geometric mean of the growth rate $\bar{\mu}$ differed between the two figures: $\bar{\mu}=0.4$ in (A) and $\bar{\mu}=0.8$ in (B). For each of the three distribution families, 
we generated several variants with the same mean ($\mathbb{E}[\Delta\tau]=100$) but different variances and ran the simulations. Uniform distributions were sampled from $U(100-d,100+d)$; for $d=10,30,50,70$ and $90$. Normal distributions were sampled from $\mathcal{N}(100,\sigma^2)$; for $\sigma=3,9,15$ and $27$. Gamma distributions were sampled from $\Gamma(k,100/k)$; for $k=2,4,5,10$ and $20$.
At each nutrient supply event, one of the two nutrient types (A or B) is chosen at random.
For each distribution, we performed 20 simulations and averaged the results.
The error bars indicate standard deviation. All the other parameters were identical to those used in Fig~\ref{fig:avePlot}.
\fi
\paragraph*{S6 Fig.}\label{fig:S6fig}
\ifshowfig
\mbox{}\par
\begin{figure}[H]
\centering
\includegraphics[]{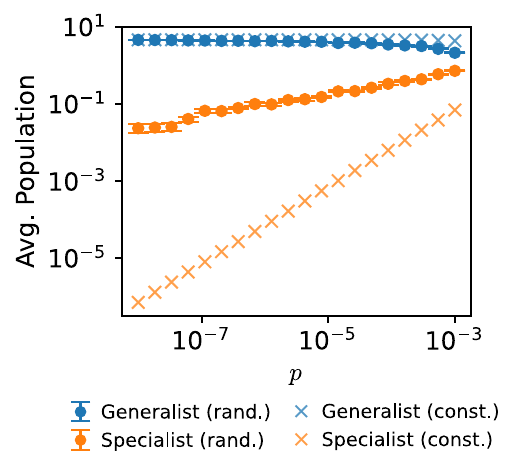}
\caption*{
\red{{\bf Effect of phenotypic switching rate on temporally averaged populations under generalist-dominant conditions. }We systematically varied the phenotypic switching rate $p$ and plotted the temporally averaged populations with $\bar{\mu}=0.8$. Crosses (\scalebox{0.8}{$\times$}) denote results with constant nutrient supply intervals ($\Delta\tau_{\rm{const.}}=100$), whereas circles (\textbullet) denote those with gamma-distributed intervals ($\Delta\tau\sim\Gamma(2,50)$). Populations were averaged from $t=2.5\times10^5$ to $t=3.0\times10^5$ across 10 simulations; error bars indicate standard error. At each nutrient supply event, one of the two nutrient types (A or B) is chosen at random. The other settings were the same as those in Fig~\ref{fig:avePlot}.
}}
\end{figure}
\else
\red{\textbf{Effect of phenotypic switching rate on temporally averaged populations under generalist-dominant conditions.} We systematically varied the phenotypic switching rate $p$ and plotted the temporally averaged populations with $\bar{\mu}=0.8$. Crosses (\scalebox{0.8}{$\times$}) denote results with constant nutrient supply intervals ($\Delta\tau_{\rm{const.}}=100$), whereas circles (\textbullet) denote those with gamma-distributed intervals ($\Delta\tau\sim\Gamma(2,50)$). Populations were averaged from $t=2.5\times10^5$ to $t=3.0\times10^5$ across 10 simulations; error bars indicate standard error. At each nutrient supply event, one of the two nutrient types (A or B) is chosen at random. The other settings were the same as those in Fig~\ref{fig:avePlot}.
}
\fi
\paragraph*{S7 Fig.}\label{fig:S7fig}
\ifshowfig
\mbox{}\par
\begin{figure}[H]
\centering
\includegraphics[]{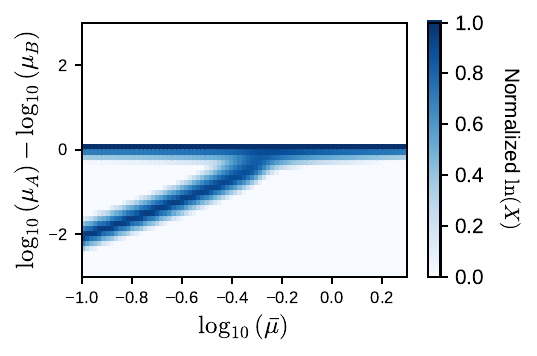}
\caption*{\red{{\bf Simulation results with an asymmetric phenotype pool. }This analysis follows the same procedure as in Fig~\ref{fig:e3}A, but with a restricted set of candidate phenotypes. From the original $N=51$ phenotypes, we excluded all strategies satisfying $\mu_{i,A}>\mu_{i,B}$. The plotting method and parameters are identical to those in Fig~\ref{fig:e3}A.}}
\end{figure}
\else
\red{\textbf{ Simulation results with an asymmetric phenotype pool. }This analysis follows the same procedure as in Fig~\ref{fig:e3}A, but with a restricted set of candidate phenotypes. From the original $N=51$ phenotypes, we excluded all strategies satisfying $\mu_{i,A}>\mu_{i,B}$. The plotting method and parameters are identical to those in Fig~\ref{fig:e3}A.}
\fi

\paragraph*{S1 Text.}
\label{sec:SI}
\textbf{Details of analytical calculations.}

\nolinenumbers
\section*{Acknowledgements}
This work was supported by JSPS KAKENHI (Grant Numbers JP22H05403 and JP25H01390 to Y.H.; 22K21344 and 23K27164 to C. F.) and JST PRESTO (Grant Number JPMJPR25K9 to Y.H.).


\end{document}